\documentstyle[prl,aps,epsf,axodraw]{revtex}
\def\bbox#1{\mbox{\boldmath$#1$}}
\begin{document}
\title{Asymmetry of the natural line profile for the hydrogen atom.}

\author{L. N.~Labzowsky${}^{1,2}$, D. A.Solovyev${}^2$, 
        G.~Plunien${}^3$, and G.~Soff${}^3$}
\address{${}^1$ {Max-Planck-Institut f\"ur Physik Komplexer Systeme,
         N\"othnitzer Stra\"se 38,D-01187 Dresden, Germany}}
\address{${}^2$ {Institute of Physics,
             St.~Petersburg State University 198904,
             Petershof, St.~Petersburg, Russia}}
\address{${}^3$ {Institut f\"ur Theoretische Physik,
             Technische Universit\"at Dresden,
             Mommsenstra{\ss}e 13, D-01062, Dresden, Germany}}
\date{\today}
\maketitle
\begin{abstract}
   The asymmetry of the natural line profile for transitions
   in hydrogen-like atoms is evaluated within a QED framework.
   For the Lyman-alpha $1s-2p$ absorption transition in neutral
   hydrogen this asymmetry results in an additional energy
   shift of 2.929856 Hz. For the $2s_{1/2}-2p_{3/2}$ transition 
   it amounts to -1.512674 Hz. 
   As a new feature this correction turns out to be process dependent. 
   The quoted numbers refer to the Compton-scattering process.
   \end{abstract}


\bigskip
The problem of the natural line profile in atomic physics was
considered first in terms of quantum mechanics by Weisskopf
and Wigner \cite{weisskopf30}. Within the framework 
of modern QED it was first formulated
for one-electron atoms by Low \cite{low51}. In \cite{low51} the 
occurance of the Lorentz profile in the resonance approximation 
was described and the non-resonant corrections were estimated. 
Later the line profile QED theory has been modified also
for many-electron atoms \cite{labzowsky83} (see also 
\cite{labzowsky93b})
 and has been applied to
the theory of overlapping resonances in two-electron highly
charged ions  \cite{karasiev92}. 
Within this formalism  
corrections to the energy levels, e.g. reference state QED 
corrections, have been evaluated \cite{labzowsky93karasiev}.

  One of the important consequences of the line profile
  theory is the occurance of non-resonant corrections
  \cite{low51}. These corrections indicate the limit up to which the
  concept of the energy of an excited atomic state has a
  physical meaning - that is the resonance approximation. The exact
  theoretical value for the energy of an excited state
  defined, e.g., by the Green function pole, can be compared
  directly with measurable quantities only within the resonance
  approximation, for which the line profile is described by the two
  parameters: energy $E$ and width $\Gamma$. Beyond this approximation the
  evaluation of $E$ and $\Gamma$ should be replaced by the evaluation of
  the line profile for the particular process. If
  the distortion of the Lorentz profile is still small one can
  formally consider the non-resonant correction as a correction
  to the energy shift. Unlike all other energy corrections, this
  correction depends on the particular process under consideration
  which has been employed for the measurement of 
  the energy difference. The non-resonant (NR)
  corrections were considered for H-like ions of phosphorus ($Z=15$) and uranium ($Z=92$)
  in \cite{labzowsky94}, \cite{labzowsky97goidenko}. While for uranium the NR correction 
  was found to be negligible, its value was comparable with the
  experimental inaccuracy in the case of phosphorus.

  In this paper we demonstrate that the concept of a transition energy depends 
  on the measurement process. For this reason 
  we evaluate the NR corrections for the neutral
  hydrogen atom. We consider the process of the resonance Compton
  scattering as a standard procedure for the determination of the
  energy levels. For this process the parametric estimate of the NR
  correction can be expressed as \cite{low51} (in relativistic units).
  \begin{equation}
  \delta=Cm\alpha^2(\alpha Z)^6
  \label{eq:corr}
  \end{equation}
  where $C$ is some numerical factor, $\alpha$ is the fine structure
  constant, $Z$ is the nuclear charge number.

  The recent QED calculations for low-$Z$ H-like atoms
  incorporate corrections of the order of $m\alpha^2(\alpha Z)^5$
\cite{pachucki} - \cite{eides}; corrections of order $m\alpha^2(\alpha Z)^6\ln^3(\alpha Z)$
  are also included \cite{karshenboim}, \cite{yerokhin}. For low-$Z$ accurate 
  Lamb-shift calculations to all orders in $Z\alpha$ have been performed
  recently \cite{jentschura99}, \cite{jentschura01}.

  Thus in principle the next order corrections to the energy
  levels should include NR corrections and will depend on the
  process of measurement under consideration. However the
  numerical factor $C$ in Eq. (1) appears to be quite
  small: $10^{-3}$ for the Lyman-alpha transition (see below). 
  There are no direct measurements of the Lyman-$\alpha$ transition
  frequency with an accuracy required in order to observe the correction
  Eq. (\ref{eq:corr}). 
  Modern experimental techniques employed in Lamb-shift measurements 
  are based on two-photon resonances, 
  e.g. for the transition 2s -1s [17] - [19]. Although
  the theoretical evaluation of the NR corrections to the two-photon resonances 
  is more involved, we can state that the corresponding NR corrections 
  will be of the same order of magnitude as the one considered here.

  Consider the process of photon scattering on a one-electron
  atom. This process is described by the two Feynman diagrams of Fig. 1.
  Resonance scattering implies that the frequency of the
  initial photon $\omega$ is close to the energy difference
  $\omega=E_{A'}-E_B$ where $A'$ is some excited atomic state,
  $B$ is the initial state. Within
  the resonance approximation we retain only the term $n=A'$ in
  the sum over intermediate states in the amplitude, corresponding
  to Fig. 1a. The amplitude, corresponding to Fig. 1b, has a
  non-resonant character. Separating out the resonant term we express
  the amplitude corresponding to Fig. 1 in the form
\begin{eqnarray}
  U^{(2)}_{AB}(\omega jm\lambda;\omega'j'm'\lambda')
  &=&
  e^2\left(
  \frac{({\bbox{\gamma}\bbox{A}^*_{\omega'j'm'\lambda'})_{AA'}}
  (\bbox{\gamma}\bbox{A}_{\omega
  jm\lambda})_{A'B}}{E_{A'}-E_B-\omega} 
\right.
\nonumber \\
&&
\left.
  +\sum_{n\neq A'}\,\frac{({\bbox{\gamma}\bbox{A}^*_{\omega'j'm'\lambda'})_{An}}
  (\bbox{\gamma}\bbox{A}_{\omega
  jm\lambda})_{nB}}{E_n-E_B-\omega}
\right.
\nonumber\\
&&
\left.
  +\sum_n\,\frac{({\bbox{\gamma}\bbox{A}_{\omega jm\lambda})_{An}}
  (\bbox{\gamma}\bbox{A}^*_{\omega'
  j'm'\lambda'})_{nB}}{E_n-E_A+\omega}\right)
\end{eqnarray}
where $\bbox{\gamma}$ are Dirac matrices, $\bbox{A}_{\omega
jm\lambda}$ is the vector potential of the electromagnetic
field (photon wave function) and $E_n$ are the one-electron
energies.

The second and third term in Eq. (2) represent the non-resonant
corrections to the scattering amplitude. 

The Lorentz line profile arises when we sum up all the electron
self-energy insertions in the internal electron line in Fig. 1a
within the resonance approximation \cite{low51}. 
After the summation of the arising
geometric progression one finds
\begin{eqnarray}
  U^{(2)}_{AB(A')} &=&
  e^2\,
  \frac{({\bbox{\gamma}\bbox{A}^*_{\omega'j'm'\lambda'})_{AA'}}
  (\bbox{\gamma}\bbox{A}_{\omega
  jm\lambda})_{A'B}}{E_{A'}-E_B+(\hat{\Sigma}_R(\omega+E_A))_{A'A'}-\omega}
  \, ,
\end{eqnarray}
where $\hat{\Sigma}_R(\omega+E_A)$ is the renormalized electron
self-energy operator.

In the resonance approximation we can 
substitute $\omega+E_A=E_{A'}$ in
the denominator of Eq. (3). The real part of the matrix element
$\hat{\Sigma}_R(\omega+E_A))_{A'A'}$ yields the lowest-order
contribution to the Lamb shift while the imaginary part which is
finite and not subject to renormalization indicates the total
radiative (single-quantum) width of the level $A'$:
\begin{eqnarray}
\hat{\Sigma}_R(\omega+E_A))_{A'A'}=L^{{\rm SE}}_{A'}-
\frac{1}{2} \Gamma_{A'}\, .
\end{eqnarray}

An additional contribution to the lowest-order Lamb shift
$L^{{\rm VP}}_{A'}$ originates from the vacuum polarization graph. 
However, this graph gives no contribution to the width \cite{labzowsky93b}.

Taking the square modulus of the amplitude Eq. (2), integrating over
the directions of the absorbed and emitted photons and summing
over polarizations we obtain the Lorentz profile for the
absorption probability
\begin{eqnarray}
   dW(\omega)=\frac{1}{2\pi}
\frac{\Gamma_{A'A}\,d\omega}{(\omega^0_{A'A}+L^{{\rm SE}}_{A'}-\omega)^2+1/4\,\,\Gamma_{A'}}\, .
\end{eqnarray}
Here $dW(\omega)$ is the probability for photon absorption in
the frequency interval
$[\omega,\omega+d\omega]$, $\omega^0_{A'A}=E_{A'}-E_A$ and
$\Gamma_{A'A}$ is the partial width of the level $A'$, connected
with the transition $A'\rightarrow A$. The inclusion of the Lamb
shift $L_{A'}$ corresponding to the initial state $A$ into the
Lorentz profile (5) can be accomplished, if necessary, by the
methods developed in \cite{labzowsky83}, \cite{labzowsky93b}.
 The line profile for the emission
process $A'\rightarrow A$ is described again by Eq. (5).

In the non-resonant terms in Eq. (2) we can substitute
$\omega=E_{A'}-E_B$. Then we arrive at the expression (we omitted
 the Lamb shift $L_{A'}$, which is not essential for our purposes) 
\begin{eqnarray}
  U_{AB}(\omega jm\lambda;\omega'j'm'\lambda')
  &=&
  e^2\left(
  \frac{({\bbox{\gamma}\bbox{A}^*_{\omega'j'm'\lambda'})_{AA'}}
  (\bbox{\gamma}\bbox{A}_{\omega
  jm\lambda})_{A'B}}{E_{A'}-E_B-\omega-i\Gamma_{A'}/2}
\right.
\nonumber\\
&&
\left.
  +\sum_{n\neq A'}\,\frac{({\bbox{\gamma}\bbox{A}^*_{\omega'j'm'\lambda'})_{An}}
  (\bbox{\gamma}\bbox{A}_{\omega
  jm\lambda})_{nB}}{E_n-E_{A'}}
\right.
\nonumber\\
&&
\left.
  +\sum_n\,\frac{({\bbox{\gamma}\bbox{A}_{\omega jm\lambda})_{An}}
  (\bbox{\gamma}\bbox{A}^*_{\omega'
  j'm'\lambda'})_{nB}}{E_n+E_{A'}-E_A-E_B}\right)\, .
\end{eqnarray}

The differential cross section of the process is
\begin{eqnarray}
d\sigma_{AB}(\omega jm\lambda;\omega'j'm'\lambda')=2\pi|U^{(2)}_{AB}(\omega
jm\lambda;\omega'j'm'\lambda')|^2\delta(E_A-E_B+\omega{'}-\omega)d\omega
d\omega{'}
\end{eqnarray}

The one-electron states $A, A', B$ depend on the usual set of
quantum numbers $A\equiv{n_Al_Aj_Am_A}$, where $n_A$ is the principal
quantum number, $l_A$ is the orbital angular momentum quantum number which
determines the parity, $j_A,m_A$ are the total angular momentum
and its projection. Integration over $\omega'$, summation over
$j'm'\lambda',m_A$ and averaging over $m,m_B$ yields :
\begin{eqnarray}
\sigma_{AB}(\omega jm\lambda) &=& \nonumber \\ 
& &\hspace{-1.5cm}\frac{2\pi}{(2j_B+1)(2j+1)}\sum_{j'\lambda'}\sum_{mm'}\sum_{m_Am_B}
\left|U^{(2)}_{AB}\left(\omega jm\lambda;
(\omega+E_B-E_A)j'm'\lambda'\right)\right|^2d\omega\, .
\end{eqnarray}

Substituting Eq. (6) in Eq. (8) and omitting the
terms, quadratically dependent on the non-resonant
contributions, we obtain
\begin{eqnarray}
\sigma_{AB}=\sigma^{(0)}_{AB}+\sigma^{(1)}_{AB}\, .
\end{eqnarray}

The first term in Eq. (9) leads to the usual Lorentz line profile
for the process under consideration
\begin{eqnarray}
\sigma^{(0)}_{AB}(\omega
jm\lambda)=\frac{1}{2\pi}\frac{2j_{A'}+1}{2j+1}
\frac{\Gamma_{AA'}\,W_{BA'}(j\lambda)}{(E_{A'}-E_B-\omega)^2+\Gamma^2_{A'}/4}
\end{eqnarray}
where $W_{BA'}$ is the transition probability $B\rightarrow
{A'}$
\begin{eqnarray}
W_{BA'}=\frac{2\pi}{2j_B+1}\sum_{mm_Bm_{A'}}|(\bbox{\gamma} \bbox{A^*}_{\omega jm\lambda})_{BA'}|^2
\end{eqnarray}
and $\Gamma_{AA'}$ is the partial width of the level
$A'$, connected with the transition ${A'}\rightarrow A$
\begin{eqnarray}
\Gamma_{AA'}=\sum_{j'\lambda'}W_{AA'}(j'\lambda')\, .
\end{eqnarray}

It is assumed that the level $B$ is stable ($\Gamma_B=0$) or
metastable ($\Gamma_B\ll{\Gamma_{A'}}$).

The term $\sigma^{(1)}_{AB}$  that represents the interference
between the resonant and non-resonant contributions to the
amplitude is determined by
\begin{eqnarray}
\sigma^{(1)}_{AB}(\omega
j\lambda)
&=&
\frac{1}{2\pi}\frac{2j_{A'}+1}{2j+1}\,{\rm Re}\left[{\sum_{n\neq{A'}}
\frac{\Gamma_{AA;A'n}\,W_{BB;nA'}(j\lambda)}{(E_{A'}-E_B-\omega- 
i\Gamma_{A'}/2)^*(E_n-E_{A'})}}
\right.
\nonumber\\
&&
\left.
+\sum_n\frac{\Gamma_{An;A'B}W_{nB;AA'}(j\lambda)}{(E_{A'}-E_B-
\omega- i\Gamma_{A'}/2)^*(E_n+E_{A'}-E_A-E_B)}\right]d\omega
\end{eqnarray}
where $W_{BB;nA'}(j\lambda)$ and $W_{nB;AA'}(j\lambda)$ are the
"mixed" transition probabilities
\begin{eqnarray}
W_{BB;nA'}(j\lambda) &=& \frac{2\pi}{2j_{A'}+1}\sum_{mm_Bm_{A'}m_n}(\bbox{\gamma}\bbox{A}_{\omega
jm\lambda})_{nB}(\bbox{\gamma}\bbox{A^*}_{\omega jm\lambda})_{BA'}\, ,
\end{eqnarray}
\begin{eqnarray}
W_{nB;AA'}(j\lambda) &=& \frac{2\pi}{2j_{A'}+1}\sum_{mm_Am_Bm_{A'}m_n}(\bbox{\gamma}\bbox{A}_{\omega
jm\lambda})_{An}(\bbox{\gamma}\bbox{A^*}_{\omega jm\lambda})_{BA'}\, ,
\end{eqnarray}
and $\Gamma_{AB;CD}$ is the "mixed" partial width
\begin{eqnarray}
\Gamma_{AB;CD}=\sum_{j'\lambda'}W_{AB;CD}(j'\lambda')\, .
\end{eqnarray}

It is important to emphasize that in the sum over $n$ in Eq. (13) only
the states $n$ with the same symmetry (i.e. with the same
$j_{A'} \lambda_{A'}$) survive. Thus we keep the same averaging factor
$(2j_{A'}+1)^{-1}$ in Eqs. (14), (15). Note also that for the fixed
transition $B\rightarrow{A'}$ the type of the absorbed photon
($j \lambda$) will be also fixed.

We assume that the standard way of measuring the resonance
frequency is employed which is connected with the determination of
the maximum in the probability distribution for the given process.
In the pure resonance case the maximum condition
\begin{eqnarray}
\frac{d\sigma^{(0)}_{AB}(\omega)}{d\omega}=0
\end{eqnarray}
corresponds to 
the resonance frequency value $\omega_{{\rm max}}=E_{A'}-E_B$. If we take
into account the correction $\sigma^{(1)}_{AB}(\omega)$, the
result will be different
\begin{eqnarray}
\frac{d}{d\omega}(\sigma^{(0)}_{AB}(\omega)+\sigma^{(1)}_{AB}(\omega))=0
\end{eqnarray}
with 
\begin{eqnarray}
\omega_{{\rm max}}=E_{A'}-E_B+\delta
\end{eqnarray}
and 
\begin{eqnarray}
\delta&=&\frac{1}{4}\frac{\Gamma^2_{A'}}{\Gamma_{AA'}W_{BA'}}Re\left[\sum_{n\neq{A'}}\frac{\Gamma_{AA;A'n}W_{BB;nA'}(j\lambda)}{E_n-E_{A'}}
\right.
\nonumber\\
&&
\left.
+\sum_{n}\frac{\Gamma_{An;A'B}W_{nB;AA'}(j\lambda)}{E_n+E_{A'}-E_A-E_B}\right]\, .
\end{eqnarray}

Thus the value of $\omega_{{\rm max}}$ for the probability
distribution in the photon scattering process cannot be compared
directly with the energy difference between the two levels. 
The process-dependent
non-resonant correction $\delta$ should be taken into
account. This result holds also if we include any QED corrections
in $E_{A'},E_B$. The order of magnitude of the correction $\delta$
follows from the standard estimates for the allowed transition
probabilities (also the "mixed" ones) $m\alpha(\alpha Z)^4$ and
the transition energies $m(\alpha Z)^2$:
\begin{eqnarray}
\delta\approx\frac{[m\alpha(\alpha Z)^4]^2}{m(\alpha
Z)^2}=m\alpha^2(\alpha Z)^6
\end{eqnarray}

These orders of magnitude are the same whether or not the level
$E_{A'}$ is removed from $E_B$ by a fine (hyperfine) structure
splitting \cite{low51}.

\bigskip
   Consider now the neutral hydrogen atom in the non-relativistic
 approximation. We put $B=A=1s$, $A'=2p$. Then the line
 profile will correspond to the Lyman-$\alpha$ transition. In this case
 $\Gamma_{A'}=\Gamma_{AA'}=W_{BA'}$ and the matrix elements of
 the electron-photon interaction are
\begin{eqnarray}
(\bbox{\gamma}\bbox{A}^*_{\omega jm\lambda})_{BA'}=
(\bbox{\gamma}\bbox{A}_{\omega jm\lambda})_{A'B}\rightarrow(U_{1m})_{1s1s}\, ,
\end{eqnarray}
where
\begin{eqnarray}
U_{1m}=\frac{4}{3}\alpha^{3/2}\pi^{1/2}\omega^{3/2}_0rY_{1m}
\end{eqnarray}
and $\omega_0=E_{2p}-E_{1s}$. In Eq. (23) and below we use atomic units.

Then the first term of the correction $\delta$ can be written as
\begin{eqnarray}
\delta_1 &=& \frac{1}{4}\sum_{mm'}(U_{1m})^*_{1s2p_{1/2}}(U_{1m'})_{1s2p{1/2}}
\nonumber \\
&& \times \int\int
d\bbox{r_1}\bbox{r_2}\,\psi^*_{1s}(\bbox{r_1})U_{1m}(\bbox{r_1})\tilde{G}_{E_{A'}}(\bbox{r_1};\bbox{r_2})U^*_{1m'}(\bbox{r_2})\psi_{1s}(\bbox{r_2})
\end{eqnarray}
where $\tilde{G}_{E_{A'}}(\bbox{r_1};\bbox{r_2})$ is the
non-relativistic "modified" Coulomb Green function and $\psi_{1s}$
is the Schr\"odinger wave function. For the angular
integration we use the partial wave expansion for
$\tilde{G}_{E_{A'}}$
\begin{eqnarray}
\tilde{G}_{E_{A'}}(\bbox{r_1};\bbox{r_2})=
\sum_{lM}\tilde{G}^l_{E_{A'}}(r_1;r_2)\,Y^*_{lM}(\Omega_1)Y_{lM}(\Omega_2)\, ,
\end{eqnarray}
where $Y_{lM}(\Omega)$ are the spherical harmonics. The
angular integration and the summation over the angular momentum
projections yields :
\begin{eqnarray}
\delta_1=\frac{1}{2}\,\frac{\alpha^6}{3^7}\int_0^{\infty} \int_0^{\infty}dr_1dr_2\,
r^3_1r^3_2\,\psi^*_{1s}(r_1) \tilde{G}^1_{E_{A'}}(r_1;r_2)\psi_{1s}(r_2)\, .
\end{eqnarray}

We evaluate Eq. (26)  
with the Sturmian expansion for $\tilde{G}^1_{E_{A'}}(r_1;r_2)$
that we take from \cite{rapoport}:
\begin{eqnarray}
\tilde{G}^l_{E_{A'}}(r_1;r_2)&=&\frac{1}{2}\sum^{\infty}_{m=l+1,m\neq n}\frac{m^4}{m-n}R_{ml}(r_1)R_{ml}(r_2)
\nonumber\\
&&
+4R_{nl}(r_2)\left\{5/4R_{nl}(r_1)+r_1\frac{d}{dr_1}R_{nl}(r_1)\right\}
\nonumber\\
&&
+ 4R_{nl}(r_1)\left\{5/4R_{nl}(r_2)+r_2\frac{d}{dr_2}R_{nl}(r_2)\right\}\, .
\end{eqnarray}
where $R_{nl}$ are the radial Coulomb wave functions. Insertion
of Eq. (27) in Eq. (26) and integration over $r_1, r_2$ for
$E_{A'}=2p$ results in
\begin{eqnarray}
\delta_1=\frac{\alpha^6}{3}\left(\frac{2}{3}\right)^{16}
\left(\sum^\infty_{m=3}\frac{(m+1)!}{(m-2)(m-2)!}\,
\phantom{.}^{\phantom{2}}_2F_1^2(2-m,5;4;2/3)+7/2\right)\, ,
\end{eqnarray}
where $\phantom{.}^{\phantom{2}}_2F_1$ denotes a hypergeometric function.
The expansion in Eq. (28) converges very fast and for $m=10$ it
gives an error less than $10^{-6}$. We obtain
\begin{eqnarray}
\delta_1 = 2.127209\cdot 10^{-3} \alpha^6 = 2.1168998 \, {\rm Hz}\, .
\end{eqnarray}

The second term of the correction $\delta$ can be written again as
Eq. (26) but with the "normal" Coulomb Green function 
$G_{E_A+E_B-E_{A'}}(\bbox{r_1};\bbox{r_2})=G_{-7/8}(\bbox{r_1};\bbox{r_2})$.
In this case it is
convenient to use for the radial Green function $G^l_E(r_1;r_2)$
the representation \cite{rapoport}
\begin{eqnarray}
G^l_E(r_1;r_2)=\frac{Z}{\nu}\sum^\infty_{m=2}\frac{m^4}{m-\nu}R_{ml}(2r_1/\nu)R_{ml}(2r_2/\nu)
\end{eqnarray}
where $\nu=Z/\sqrt{-2E}=2/\sqrt{7}$. Now we obtain
\begin{eqnarray}
\delta_2&=&4\alpha^6\frac{ \nu^7}{
(\nu+1)^{10}}\left(\frac{2}{3}\right)^7\sum^{\infty}_{m=2}
\frac{(m+1)m(m-1)}{m-\nu}\phantom{.}^{\phantom{2}}_2F_1^2(2-m,5;4;2/(\nu+1))\, .
\end{eqnarray}
Retaining only six terms of the expansion (31) yields an
accuracy of $10^{-6}$. Then
\begin{eqnarray}
\delta_2 = 0.821625\cdot 10^{-3} \alpha^6=0.81764337 \, {\rm Hz}
\end{eqnarray}
and finally
\begin{eqnarray}
\delta^{(2p)}_{1s,1s} = \delta_1+\delta_2 = 2.929856 \, {\rm Hz}\, .
\end{eqnarray}

It should be mentioned, that the Lyman-$\alpha$
resonance consists of two peaks in the scattering experiment, 
corresponding to the two
fine-structure components. In the non-relativistic approximation
the distortion of these two peaks is equal and defined by
Eq. (33). 

We made also an analogous calculations for the transition 
$2s_{1/2}\rightarrow 2p_{3/2}\rightarrow
1s_{1/2}$. In this case $B=2s_{1/2}$, 
$A=1s_{1/2}$, $A'=2p_{3/2}$. The result is
\begin{eqnarray}
\delta^{(2p_{1/2})}_{2s,1s} = -\left(\frac{2}{3}\right)^{16}\alpha^6
= -1.522439\cdot 10^{-3} \alpha^6 = -1.512674 \, {\rm Hz}\, .
\end{eqnarray}

Concluding, we can state that the NR corrections are comparable
with some QED corrections to the Lamb shift recently included in
the consideration \cite{eides}. The level of accuracy of
modern experiments with neutral hydrogen for the two-photon transition
$2s - 1s$ is estimated in total to be $46$ Hz 
\cite{hieringother} and 
is approaching the magnitude of these non-resonant corrections.
The derivation of the two-photon NR corrections requires consideration
of two-photon resonance scattering on a hydrogen atom. The corresponding
expression for the amplitude will contain two energy denominators unlike as
in Eq. (2). However, only one of them will become resonant
while the other one appears as a nonresonant factor
and will not change
the order of magnitude of the total amplitude. Accordingly, the 
interference term (13) defining the resonance shift will lead to values
of the same order of magnitude as the corrections (33) and (34), 
respectively. 

\begin{center}
Acknowledgements
\end{center}

 The work of L.L. and D.S. was supported by the RFBR grant ü99-02-18526.
G.P. and G.S. acknowledge financial support from BMBF, DFG and
GSI.

\begin{center}
\begin{figure}
\begin{picture}(400,400)(0,0)
\Line(15,390)(15,210)\Text(9,300)[]{$n$}
\Line(13,390)(13,210)\Text(9,390)[]{$A$}
\Photon(14,370)(100,370){3}{6}\Text(9,210)[]{$B$}
\Photon(14,230)(100,230){3}{6}\Text(55,383)[]{$\omega' j'm'\lambda'$}
\Text(50,240)[]{$\omega jm\lambda$}
\Text(50,150)[]{a$)$}
\Line(220,390)(220,210)\Text(214,390)[]{$A$}
\Line(218,390)(218,210)\Text(214,210)[]{$B$}
\Photon(219,370)(305,370){3}{6}\Text(214,300)[]{$n$}
\Photon(219,230)(305,230){3}{6}\Text(265,383)[]{$\omega jm\lambda$}
\Text(270,242)[]{$\omega' j'm'\lambda'$}
\Text(265,150)[]{b$)$}
\end{picture}
%
\caption{ 
 Feynman graph, corresponding to the photon scattering on an
atomic electron. The solid double line denotes the bound
electron in the atom. The wavy lines denote the absorbed
and emitted photons. The initial, intermediate and final states of
an electron are denoted as $B, n, A$. The initial and final photon
states are $\omega j m \lambda$ and $\omega' j' m' \lambda'$
where $\omega$ is the frequency, $j m$ denote the photon angular
momentum and its projection, the quantum number $\lambda$ determines
the parity of the photon state.
}
\end{figure}
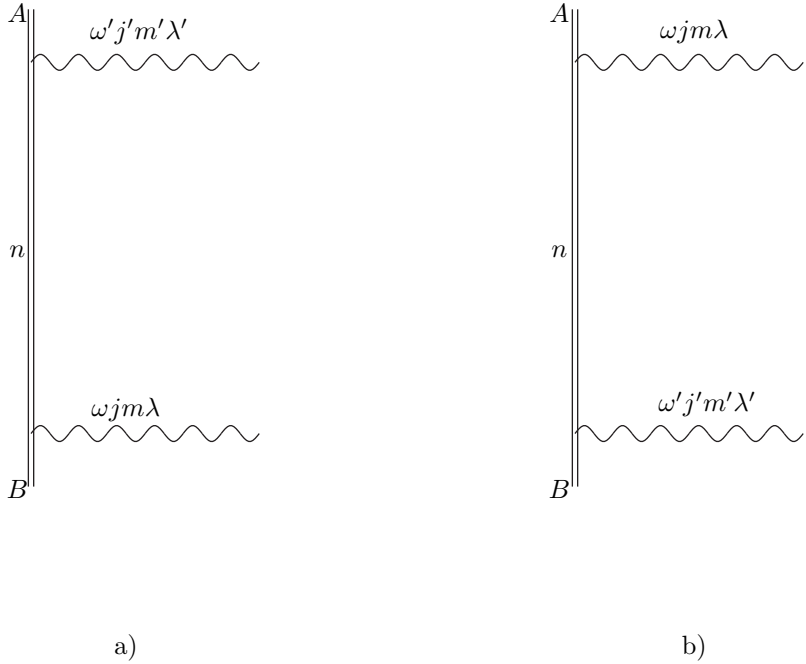
\end{center}
\end{document}